\documentclass{PoS}
\usepackage{bm,subfigure,epsfig}

\newcommand{\beq}{\begin{equation}}
\newcommand{\eeq}{\end{equation}}
\newcommand{\bea}{\begin{eqnarray}}
\newcommand{\eea}{\end{eqnarray}}

\title{The unitary Fermi gas at finite temperature: momentum distribution and contact}

\ShortTitle{The unitary Fermi gas: momentum distribution and contact}

\author{\speaker{Joaqu\'\i n E. Drut}\\
       Theoretical Division, Los Alamos National Laboratory, Los Alamos, NM 87545, USA.\\
       E-mail: \email{joaquindrut@gmail.com}}

\author{Timo A. L\"ahde\\
       Helsinki Institute of Physics and Department of Applied Physics, 
       Aalto University, FI-00076 Aalto, Espoo, Finland \\
       E-mail: \email{talahde@gmail.com}}

\author{Timour Ten\\
	Department of Physics, University of Illinois, Chicago, IL 60607, USA.\\
	Theoretical Division, Los Alamos National Laboratory, Los Alamos, NM 87545, USA. \\
	E-mail: \email{tten1@uic.edu}
	}

\abstract{
The Unitary Fermi Gas (UFG) is one of the most strongly interacting systems known to date, as it saturates the unitarity 
bound on the quantum mechanical scattering cross section. The UFG corresponds to a two-component Fermi gas in the 
limit of short interaction range and large scattering length, and is currently realized in ultracold-atom experiments via 
Feshbach resonances. While easy to define, the UFG poses a challenging quantum many-body problem, as it lacks any 
characteristic scale other than the density. As a consequence, accurate quantitative predictions of the thermodynamic 
properties of the UFG require Monte Carlo calculations. However, significant progress has also been made with purely 
analytical methods. Notably, in 2005 Tan derived a set of exact thermodynamic relations in which a universal quantity 
known as the "contact" C plays a crucial role. Recently, C has also been found to determine the prefactor of the high-
frequency power-law decay of correlators as well as the right-hand-sides of shear- and bulk viscosity sum rules. The 
contact is therefore a central piece of information on the UFG in equilibrium as well as away from equilibrium.
In this talk we describe some of the known aspects of Fermi gases at and around unitarity, show our latest Monte Carlo 
results for the contact at finite temperature, and summarize the open questions in the field, some of which we are 
starting to answer using large-scale Monte Carlo calculations by adapting methods from Lattice QCD.
}

\FullConference{The XXIX International Symposium on Lattice Field Theory - Lattice 2011\\
July 10-16, 2011\\
Squaw Valley, Lake Tahoe, California}

\begin{document}

\section{Introduction}

In recent years, our understanding of universality~\cite{Universality} in non-relativistic many-body quantum mechanics 
has increased dramatically. By universality we mean independence from the details of the interaction, in the same sense 
as in the context of second-order phase transitions, but in the absence of long-range correlations 
throughout the system (except potentially at special points in the phase diagram). Among the systems displaying this 
property, perhaps the most dramatic example is the so-called unitary Fermi gas (UFG). 
This system is a two-component Fermi gas tuned to the limit of vanishing interaction range $r^{}_0$ and large s-wave 
scattering length $a^{}_s$, and is termed "unitary" because it saturates the unitarity bound imposed on the scattering 
cross section by the unitarity of quantum mechanics. In short, the unitary gas is a resonant quantum mechanical 
many-body system.

A few years ago, the unitary limit was realized in metastable ultracold atomic clouds in various laboratories around the 
world~\cite{FirstExperiments} and it has been under intense scrutiny by the atomic, molecular and optical physics 
 (AMO) community ever since~\cite{Reviews}. Interest in the UFG transcends those areas, however, with a considerable 
amount of research being carried out within the nuclear physics community well before and after the first AMO 
experiments~\cite{NucPhys}. This, of course, is itself a manifestation of the universality of the UFG, as nuclear 
systems characteristically display short ranges and unnaturally large scattering lengths, although in a vastly different 
absolute scale than atomic clouds, the natural scale being in each case the Fermi momentum $k^{}_F$.

More recently, the limit $0 \leftarrow k^{}_F r^{}_0 \ll 1 \ll k^{}_F a^{}_s \rightarrow \infty$ has been shown to imply 
non-relativistic conformal invariance, as described in Ref.~\cite{NishidaSon}. In turn, this results in a set of 
non-trivial relations between the system in homogeneous space and in a harmonic trap, as first shown in 
Ref.~\cite{Mehen,Werner}. In a separate line of research, short-distance correlations were shown by 
Tan~\cite{Shina} and others~\cite{ZhangLeggett, BraatenPlatter} to be completely encoded in a quantity $C$, 
which Tan called the ``contact''. Specifically, we may define the contact as 
\beq
\label{Cdef}
C \equiv \lim_{k \to \infty} k^4_{} n_\sigma^{}(k),
\eeq
where $n_\sigma^{}(k)$ is the momentum distribution for spin~$\sigma$ expressed as a thermal average. 
This remarkable property stems in part from the short-range nature of 
the interaction, which implies that at resonance the many-body wavefunction is essentially that of a free gas, 
with the added boundary condition that it diverges as $1/r$ when two coordinates are set to a short distance apart 
$r$~\cite{WernerCastin}. Following the work of Tan and others, the last couple of years have seen considerable activity 
extending the analysis of short-range correlations in many-body systems to systems away from unitarity as well as 
to different dimensions~\cite{WernerCastin2} and to a growing set of thermodynamic and even hydrodynamic 
quantities~\cite{SonThompson, TaylorRanderia}. The latter, in particular, point to the fact that $C$ is relevant not only in 
equilibrium but also {\it away} from equilibrium (see Ref.~\cite{Braaten} for a comprehensive review).

In spite of experimental advances and progress from the formal and analytic points of view, the UFG remains a 
challenging many-body problem. The reason for this is that, while resonant and therefore strongly interacting, the 
UFG has as few scales as a non-interacting Fermi gas. Enhanced symmetries aside, such a lack of scales implies 
lack of small parameters to perform an expansion, such that non-perturbative numerical methods are required. Indeed, 
while we know that Tan's contact plays a crucial role in the dynamics of the UFG, the only way to determine it accurately 
and reliably is by using numerical methods such as Quantum Monte Carlo (in any of its various incarnations, in 
particular on the lattice).

The work shown here represents the first attempt to determine the contact at finite temperature in a non-perturbative
fashion. To this end, we have adapted methods from Lattice QCD, namely Hybrid Monte Carlo~\cite{DuaneGottlieb} and applied them to the calculation of the momentum distribution of the UFG. The next section outlines the main features of the algorithm and Sec.~\ref{Results} shows our results and conclusions, which were first published in Ref.~\cite{DrutLahdeTenPRL}.

\section{Algorithm \& lattice formulation}

The lattice formulation we have used for this work follows closely that of Ref.~\cite{BDM}, but differs in at least 
three notable aspects. 
Firstly, we determine the bare lattice coupling constant $g$ corresponding to the unitary regime by using 
L\"uscher's formula~\cite{Luescher} as in Ref.~\cite{LeeSchaefer}, without imposing a spherically symmetric cutoff. This procedure yields $g \simeq 5.144$ in the unitary 
limit. 
Secondly, we use the compact, continuous Hubbard-Stratonovich~\cite{HST} transformation
\bea
\label{LeeHS}
\exp\left(\tau g \,\hat n_{\uparrow i}^{} \hat n_{\downarrow i}^{}\right) &=& 
\frac{1}{2\pi} \int_{-\pi}^{\pi} d\sigma_i^{}
\left[1 + B \sin(\sigma_i^{}) \, \hat n_{\uparrow i}^{}\right]\left[1 + B \sin(\sigma_i^{}) \, \hat n_{\downarrow i}^{}\right],
\eea 
where $\sigma_i^{}$ (not to be confused with the spin projection) is the auxiliary field, with 
$B^2/2 \equiv \exp(\tau g) - 1$, and $\tau$ denotes the lattice spacing in the imaginary time direction.
We find that a time step $\tau \simeq 0.05$ is sufficiently small to render 
temporal discretization errors insignificant. The above representation (referred to as ``Type~4'' in Ref.~\cite{DeanLee}) was 
found to be superior with respect to acceptance rate, decorrelation and signal-to-noise properties than the more 
conventional unbounded and discrete forms~\cite{Hirsch}. 
Finally, the use of a continuous auxiliary field allows us to perform global updates using the Hybrid Monte Carlo (HMC) 
algorithm~\cite{DuaneGottlieb}. Our implementation of the HMC algorithm does not use pseudofermions but rather 
relies on a direct calculation of the fermion determinant in a purely spatial rather than spacetime formulation~\cite{Scalapinoetal}. In addition, 
we use Fourier acceleration to propagate states in imaginary time. This enables global updates at all temperatures and 
lattice sizes, and scales approximately as $\sim V^2_{} \log V$ (at fixed temperature) for moderate spatial lattice 
volumes $V$, to be contrasted with the $\sim V^3_{}$ scaling of approaches based on local updates. 

\section{\label{Results}Results and conclusions}

\begin{figure}[h]
\includegraphics[width=.45\columnwidth]{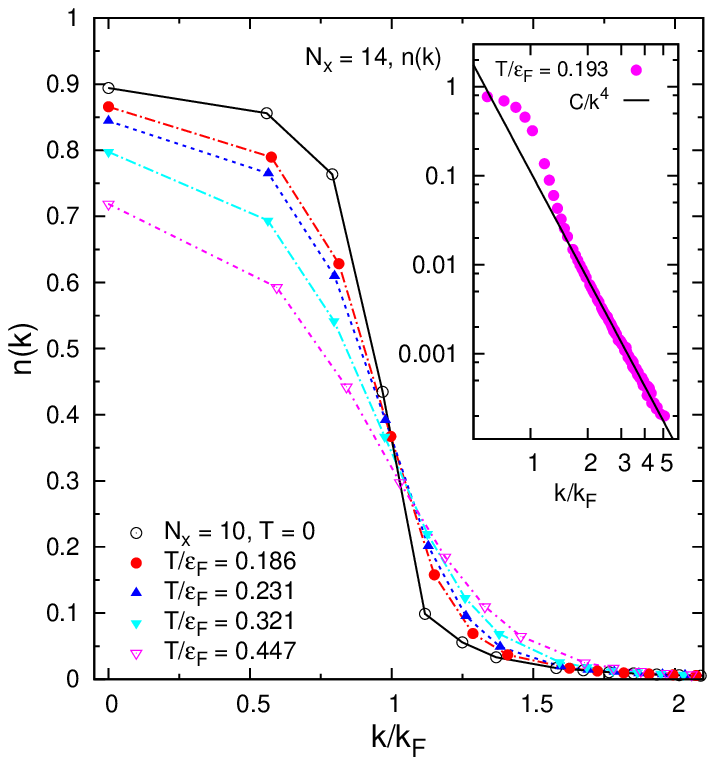}
\includegraphics[width=.45\columnwidth]{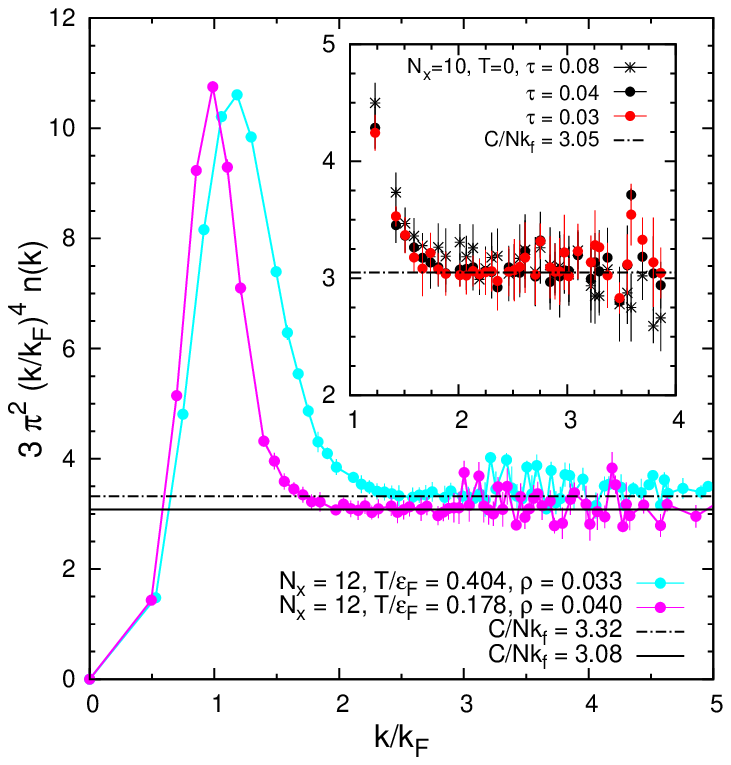}
\caption{ (Color online) Right panel: Momentum distribution $n(k)$ from QMC for $N_x^{} = 10$
as a function of $k/k_F^{}$, for various temperatures ranging 
from zero to $T/\epsilon_F^{} \simeq 0.5$. The solid lines are intended to guide the eye, and the statistical errors are
of the size of the symbols. Inset: $n(k)$ for $N_x^{} = 14$ in a log-log scale, showing the asymptotic $\sim k^{-4}_{}$ 
behavior. Left panel: Plot of $3 \pi^2 (k/k_F^{})^4_{} n(k)$ for $N_x^{} = 12$ as a function of $k/k_F^{}$ at
$T/\epsilon_F^{} = 0.178$ and $0.404$. The ``plateaux'' at large $k/k_F^{}$ give 
the intensive dimensionless quantity $C/(N k_F^{})$. At low $T/\epsilon_F^{}$, the asymptotic region is 
reached at $k/k_F^{} \simeq 2$. Inset: $N_x^{} = 10$ results at $T = 0$ showing only slight dependence on the time 
step $\tau$.}
\label{Fig:nk}
\end{figure}

\begin{figure}[t] 
\includegraphics[width=.55\columnwidth]{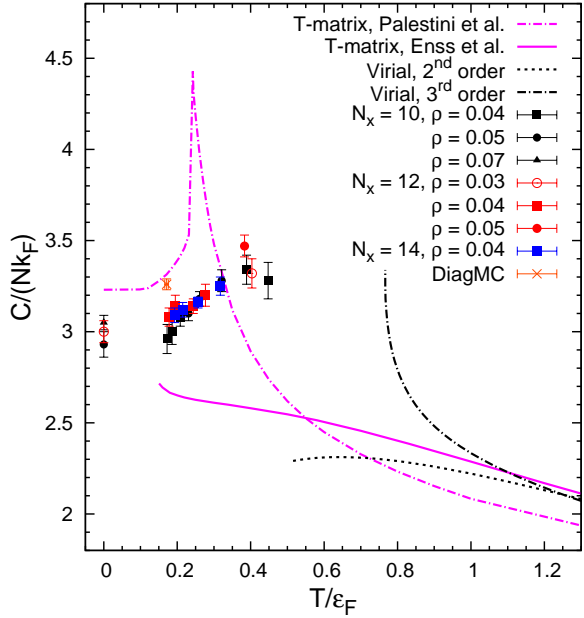}
\caption{(Color online) Summary of QMC results for $C/(N k_F^{})$ as a function of 
$T/\epsilon_F^{}$, as determined from the large $k/k_F^{}$ behavior of $n(k)$. 
The errorbars are dominated by systematics related to the residual fluctuations in the plateaux, 
as shown in the previous figure (right panel). 
Also shown are the t-matrix calculations of Ref.~\cite{Palestini, Enss}, 
the virial expansion of Ref.~\cite{Hu} and the diagrammatic Monte Carlo result of Ref.~\cite{Goulko}.} 
\label{Fig:Contact}
\end{figure}

We have performed calculations at zero as well as finite temperature, in the former case using an approach 
similar to Ref.~\cite{DeanLee}. Our main results correspond to $40-50$ particles at $N_x^{} = 10$ and 
$70-80$ particles at $N_x^{} = 12$, in addition to limited data for $N_x^{} = 14$.
In Fig.~\ref{Fig:nk} (left panel), we show the momentum distribution $n(k)$
as a function of temperature $T/\epsilon_F^{}$. 
We have computed $n(k)$ by averaging over the angular directions on the 
lattice as well as over the imaginary-time slices. In this way, we find that $\sim 200$ uncorrelated auxiliary field 
samples for each datapoint gives excellent statistics for $n(k)$. Multiplying $n(k)$ by $k^4_{}$, as plotted in 
the right panel of Fig.~\ref{Fig:nk}, we find a maximum at $k\simeq k^{}_F$ and a leveling out at high momenta, 
with the asymptotic 
regime setting in at approximately $2 k_F^{}$, at the lowest temperatures. There is no a priori reason for the asymptotic 
regime to set in at such low momenta; our work is the first to point out this fortunate situation. We study the temperature 
dependence of this ``plateau'', which allows us to determine the contact $C/(Nk_F^{})$ as a function of 
temperature. The corresponding results are given in Fig.~\ref{Fig:Contact}, together with a comparison with other 
theoretical predictions. Our results indicate that $n(k)$ follows the expected $\sim k^{-4}_{}$ dependence accurately 
up to at least $k \simeq 4 k_F^{}$, at which point the signal deteriorates, possibly due to lattice effects.

Our results show that the contact $C$ grows with temperature well beyond the superfluid phase, which is suggestive 
of a peak $C_\mathrm{max}^{} \simeq 3.4$ at $T/\epsilon_F^{} \simeq 0.4$. This scenario agrees qualitatively with 
Ref.~\cite{Yu}, as well as Ref.~\cite{Palestini}. Since $C$ measures the number of particle pairs (of both
spins) whose separation is small, the appearance of a maximum indicates an enhancement in such short-range 
correlations. We find the scale at which the $k^{-4}_{}$ law sets in (see Fig.~\ref{Fig:nk}) to be $k \simeq 2 k^{}_F$ 
at finite $T/\epsilon_F^{}$ and somewhat lower for the ground state.

In summary, we have computed the momentum distribution $n(k)$ and the contact $C/(Nk_F^{})$ for the UFG at zero 
and finite temperature, using a lattice formulation of the many-body problem, in conjunction with the HMC
algorithm. Our results represent the first fully non-perturbative calculation of $n(k)$ free of uncontrolled approximations.
We find that the contact at $T = 0$ takes the value $\simeq 2.95 \pm 0.10$ and increases as a function of 
$T/\epsilon_F^{}$ in the low- and intermediate-temperature regimes that we have explored, which is consistent with the 
phonon-dominated scenario of Ref.~\cite{Yu}. Our results complement the calculations of 
Refs.~\cite{Yu,Palestini,Enss,Hu}, and are suggestive of a maximum in $C/(Nk_F^{})$ at $T/\epsilon_F^{} \simeq 0.4$, 
which agrees qualitatively with Ref.~\cite{Palestini} but disagrees with Ref.~\cite{Enss}. While calculations at higher 
temperature $T/\epsilon_F^{} \sim 1$ are feasible, an improved understanding of the finite density effects is clearly 
called for.



\end{document}